\documentclass[slac_one]{revtex4}
\usepackage{graphicx}
\usepackage{fancyhdr}
\usepackage{epsfig}
\def\TeVc{\ifmmode {\mathrm{\ Te\kern -0.1em V}}\else
                   {\textrm{Te\kern -0.1em V}}\fi}%
\def\GeVc{\ifmmode {\mathrm{\ Ge\kern -0.1em V}}\else
                   {\textrm{Ge\kern -0.1em V}}\fi}%

\begin{document}

%Title of paper
\begin{center}
\begin{center}
\textbf{Top quark properties at ATLAS} %% Paper title goes here

% Repeat the \author .. \affiliation  etc. as needed
%
% \affiliation command applies to all authors since the last
% \affiliation command. The \affiliation command should follow the
% other information

\textbf{Dilip Jana} (\emph{on behalf of the ATLAS Collaboration}) \\
\emph{University of Oklahoma, Norman, OK. 73019, USA}                                 
\end{center}

\end{center}

\begin{abstract}
The ATLAS potential for the study of the top quark properties and physics beyond the Standard Model in the top quark sector is described. The measurements of the top quark charge, the spin and spin correlations, the Standard Model decay ($t\rightarrow bW$), rare top quark decays associated to flavour changing neutral currents ($t\rightarrow qX$ with $X$ = gluon, $Z$, photon) and $t\bar{t}$ resonances are discussed. The sensitivity of the ATLAS experiment is estimated for an expected luminosity of $1$~fb$^{-1}$ at the LHC, using full simulation Monte Carlo samples. For the Standard Model measurements the expected precision is presented. For the tests of physics beyond the Standard Model, the 5$\sigma$ discovery potential (in the presence of a signal) and the 95\% Confidence Level (C.L.) limit (in the absence of a signal) are given.
\end{abstract}

%\maketitle must follow title, authors, abstract
\maketitle

\thispagestyle{fancy}

% body of paper here - Use proper section commands
% References should be done using the \cite, \ref, and \label commands
% Put \label in argument of \section for cross-referencing
%\section{\label{}}

\section{Introduction} % Section title should be in all capitals.

The top quark was discovered in 1995 at Fermilab in pair production mode ($t\bar{t}$ events) through strong interactions[1]. Several properties of the top quark such as the mass, charge, lifetime, production cross-section and rare decay through flavour changing neutral currents (FCNC) have been explored in Fermilab, but most of these studies are limited by low statistics. Due to the high event rates at LHC (one $t\bar{t}$ event per second at a luminosity of $10^{33}$ cm$^{-2}$s$^{-1}$) at ATLAS, these top quark properties can be studied extensively at ATLAS giving the possibility to discover physics beyond the Standard Model. Due to the very short life time, the top quark decays before it has time to hadronise. But its spin properties are not washed out by hadronization, rather the top quark spin information propagates to its decay products. This unique feature allows direct top quark spin studies. The top quark spin can be reconstructed by measuring the angular distributions of its decay products in the top quark rest frame. Measurements of $W$-boson polarization complement top quark spin studies which can disentangle the origin of new physics. In the Standard Model, flavour changing neutral currents (FCNC) are strongly suppressed at the tree level due to the Glashow-Iliopoulos-Maiani mechanism. At the one loop level, small FCNC contributions are expected due to the CKM mixing matrix. The existence of $q\bar{q}$ bound states (mesons) of all other quarks encourages us to look for $t\bar{t}$ bound states. New resonances and gauge bosons strongly coupled to the top quark are expected in several theoretical models which can decay into $t\bar{t}$ pairs, leading to deviations from Standard Model $t\bar{t}$ production cross-section and top quark kinematics[2]. These new particles can reveal themselves in the $t\bar{t}$ invariant mass distribution. 

%According to the Standard Model, b-jet should be associated with a positive lepton ($\ell^{+}$), but in the case of exotic (top quark charge -4/3 e), it should be associated with a negative lepton($\ell^{-}$). So it is very important to measure top quark charge at ATLAS detector which can justify Standard Model or it may lead to new discoveries.

\section{Basic Event Selection}
We have used semileptonic ($t\bar{t}\rightarrow W W b \bar{b} \rightarrow l\nu j_1 j_2 b \bar{b}$ with $l=e,\mu$) and dileptonic ($t\bar{t}\rightarrow W W b \bar{b} \rightarrow l \nu l^{'} \nu^{'} b \bar{b}$ with $l,l^{'}=e,\mu$) decays of $t\bar{t}$ events for the top quark charge reconstruction. We have used only the semileptonic decay channel for $Wtb$ anomalous coupling, $t\bar{t}$ spin and spin correlation and $t\bar{t}$ resonance. For the semileptonic topology, we require exactly one isolated electron (muon) with $|\eta|<2.5$ and $p_{\mathrm{T}} > 25 \GeVc$ ($p_{\mathrm{T}} > 20 \GeVc$), at least 4 jets with $|\eta|<2.5$ and $p_{\mathrm{T}} > 30 \GeVc$, at least 2 jets tagged as $b$-jets and missing transverse energy above 20 \GeVc[3].  For the dileptonic topology, we require exactly two isolated electrons (muons) with $|\eta|<2.5$ and $p_{\mathrm{T}} > 25 \GeVc$ ($p_{\mathrm{T}} > 20 \GeVc$), at least 2 jets with $|\eta|<2.5$ and $p_{\mathrm{T}} > 30 \GeVc$, at least 2 jets tagged as $b$-jets and missing transverse energy above 20 \GeVc[3]. Since the final state topology for the rare top quark decays via FCNC are different from semileptonic and dileptonic topologies, we have used different selection criteria which will be described in section 2.3.    

\subsection{Top quark charge measurement}

We have presented the measurement of the top quark charge based on the reconstruction of the charge of the top quark decay products. The $W$ boson charge can be directly measured easily using its leptonic decay modes. Due to quark confinement inside hadrons, we cannot measure the $b$ quark charge directly. We have used $b$-jet charge weighting (weighted sum of all the tracks in the jet) and semileptonic $b$-decay approaches to measure $b$ quark charge. By using the weighting technique, it is possible to distinguish between the $b$-jet charges associated with leptons of opposite charges with a $5\sigma$ significance with only $0.1$~fb$^{-1}$ of data ( $1$~fb$^{-1}$ for semileptonic $b$-decay approach) which allows the Standard Model (${t\rightarrow W^{+}b}$) and exotic (${t'\rightarrow W^{-}b}$) scenarios to be distinguished. Reconstruction of the magnitude of the top quark charge seems to be possible with $\simeq 1$~fb$^{-1}$ using the weighting technique, but it is necessary to check the performance of the method with real data. The reconstructed $b$ quark and top quark charge are shown in Figure 1 with $1$~fb$^{-1}$ of simulated data. The resulting top quark charge is $Q_{t}= 0.67 \pm 0.06\;(stat) \pm 0.08\;(syst)$.      
%$Q\mathrm{_t= 0.67 \pm 0.06\;(stat) \pm 0.08\;(syst)}$.      
%($\mathrm{t\overline{t}\rightarrow W W b \overline{b} \rightarrow l \nu j_1 j_2 b \overline{b}}$

\begin{figure}[!h]
\centering
\centering\epsfig{file=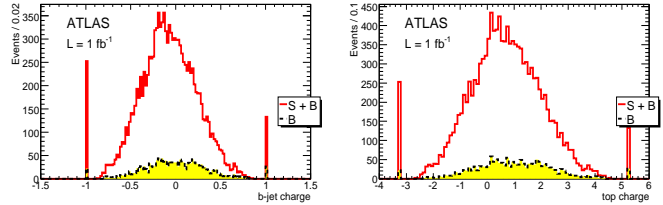, height=3cm, width=9cm, clip=}
\caption{Left: the $b$-jet charge ($Q_\mathrm{{b}}$) distribution; right: the reconstructed top quark charge ($Q\mathrm{_t}$).}
\label{bkgd_signal}
\end{figure} 

\subsection{Top quark spin and spin correlations and \emph{Wtb} anomalous couplings}

%W-boson or top quark spin information can be obtained by reconstructing the angular distributions of the daughter particles in the W-boson or top quark rest frame, respectively.

In the Standard Model, the top quarks are produced unpolarised in $t\bar{t}$ events, but their spins are correlated[4]. The production asymmetry ($A$ and $A_{\mathrm{D}}$) can be obtained from the angular distribution of the top quark decay products. In addition to the $t\bar{t}$ spin correlation, we can measure the $W$ polarization. The $W$-boson can be produced with right($F_{\mathrm{R}}$), left($F_{\mathrm{L}}$) or longitudinal polarizations($F_{\mathrm{0}}$) with $F_{\mathrm{0}} + F_{\mathrm{L}} + F_{\mathrm{R}} = 1$. The expected measurement results, using 1~fb$^{-1}$ of simulated data, are shown in Table 1. It is also possible to parameterise new physics in the $Wtb$ vertex using anomalous coupling parameters $V_{\mathrm{L}}$, $V_{\mathrm{R}}$, $g_{\mathrm{L}}$ and $g_{\mathrm{R}}$. Figure 2 shows the expected 68\% C.L. allowed regions on the $Wtb$ anomalous couplings for 1~fb$^{-1}$.

\begin{table}[htbp]
\caption{$W$-boson polarization and top quark spin correlation parameters with statistical and systematic errors.
}
\begin{center}
\begin{tabular}{|c|c|c|c|}
\hline 
$W$-boson polarization             &       $F_{\mathrm{L}}$                        &      $F_{\mathrm{0}}$                 &     $F_{\mathrm{R}}$                 \\  
                             & 0.29 $\pm$0.02 $\pm$0.03           & 0.70 $\pm$0.04 $\pm$0.02   & 0.01 $\pm$0.02 $\pm$0.02 \\ \hline\hline
 $t\bar{t}$ spin correlation &       $A$                          &       $A_{\mathrm{D}}$              &          \\
                             & 0.67 $\pm$0.17$\pm$0.18            & -0.40 $\pm$0.11 $\pm$0.09 & \\
\hline
\end{tabular}
\label{tab:pola_result}
\end{center}
\end{table}

\begin{figure}[!]
\hspace{0.022\textwidth}
\begin{minipage}[t]{.3\textwidth}
 \centering%\epsfig{file=wtb.eps,width=\linewidth}
\includegraphics[width=1.05\textwidth,angle=0]{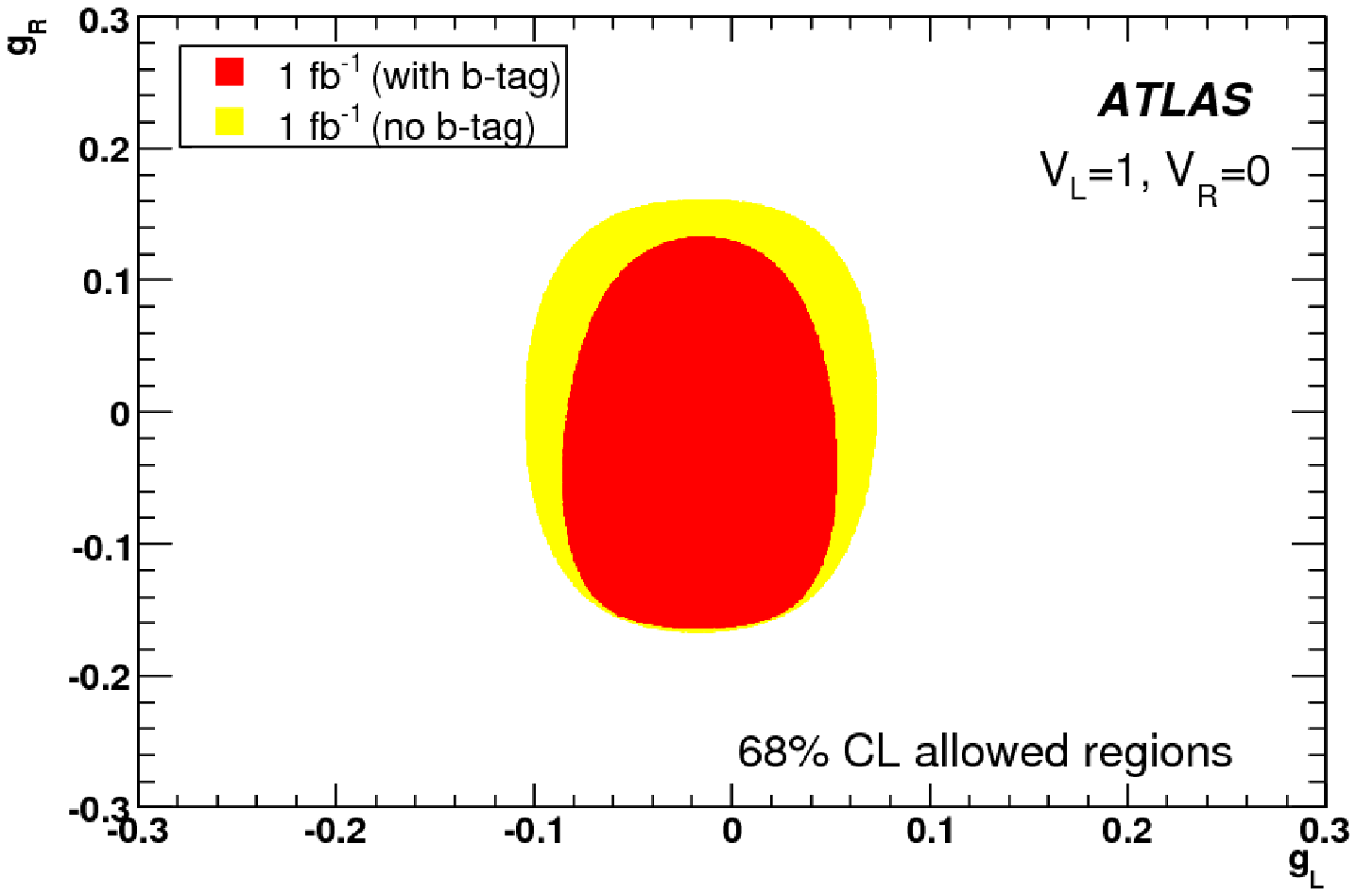}
  \caption{The expected 68\% C.L. allowed regions on the $Wtb$ anomalous couplings for $L=1$~fb$^{-1}$}
  \label{resolutionMtt}
 \end{minipage} \hfill
\begin{minipage}[t]{.25\linewidth}
 \centering%\epsfig{file=fcnc.eps,width=\linewidth}
\includegraphics[width=1.05\textwidth,angle=0]{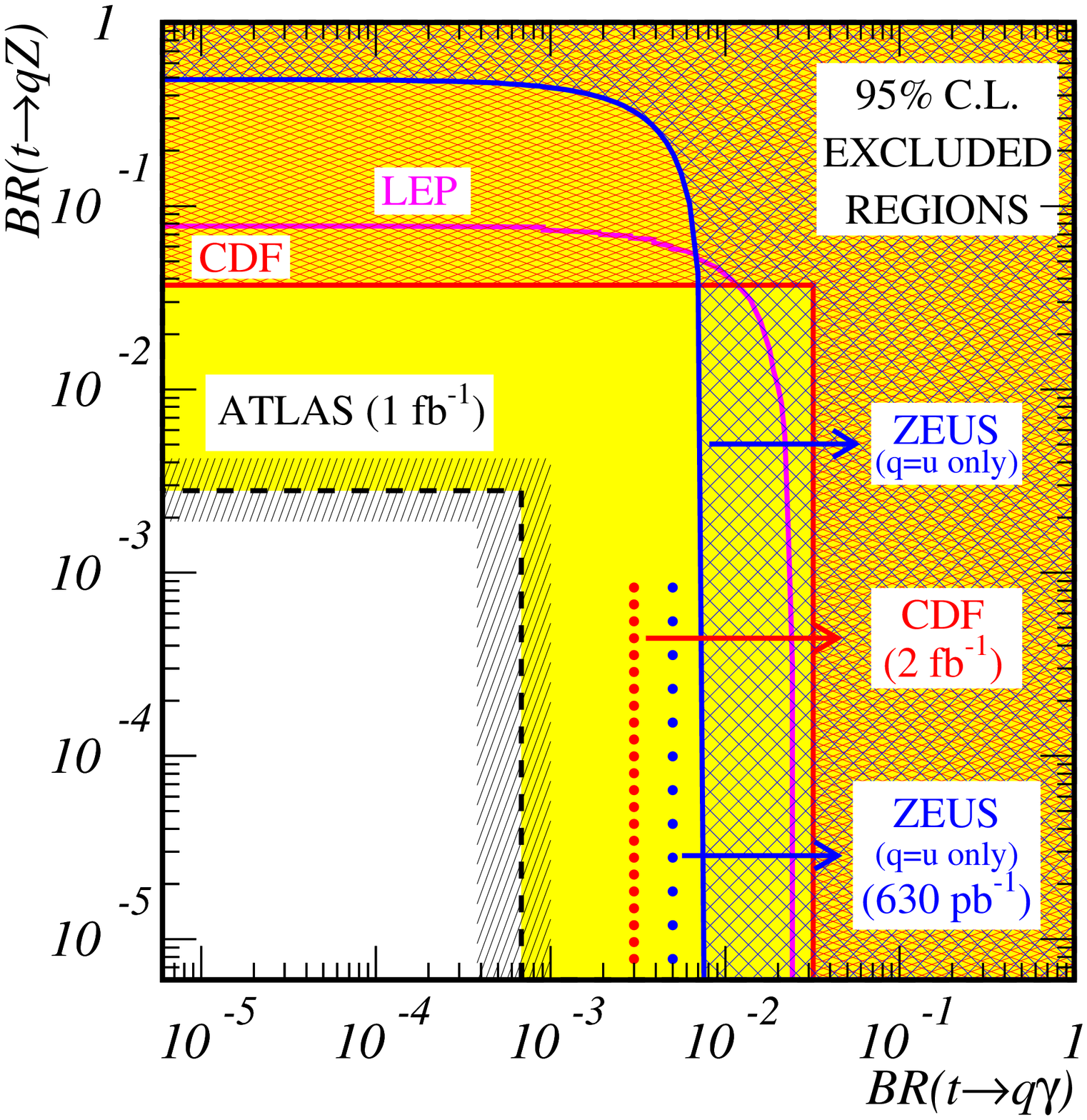}
 \caption{95\% C.L. expected limits on the $BR(t \to q\gamma)$ vs $BR(t \to qZ)$}
 
  \label{resolutionMtt}
 \end{minipage} \hfill
\hspace{0.022\textwidth}
\begin{minipage}[t]{.3\textwidth}
  \centering%\epsfig{figure=DiscoveryPotential.eps,width=\linewidth}
\includegraphics[width=1.05\textwidth,angle=0]{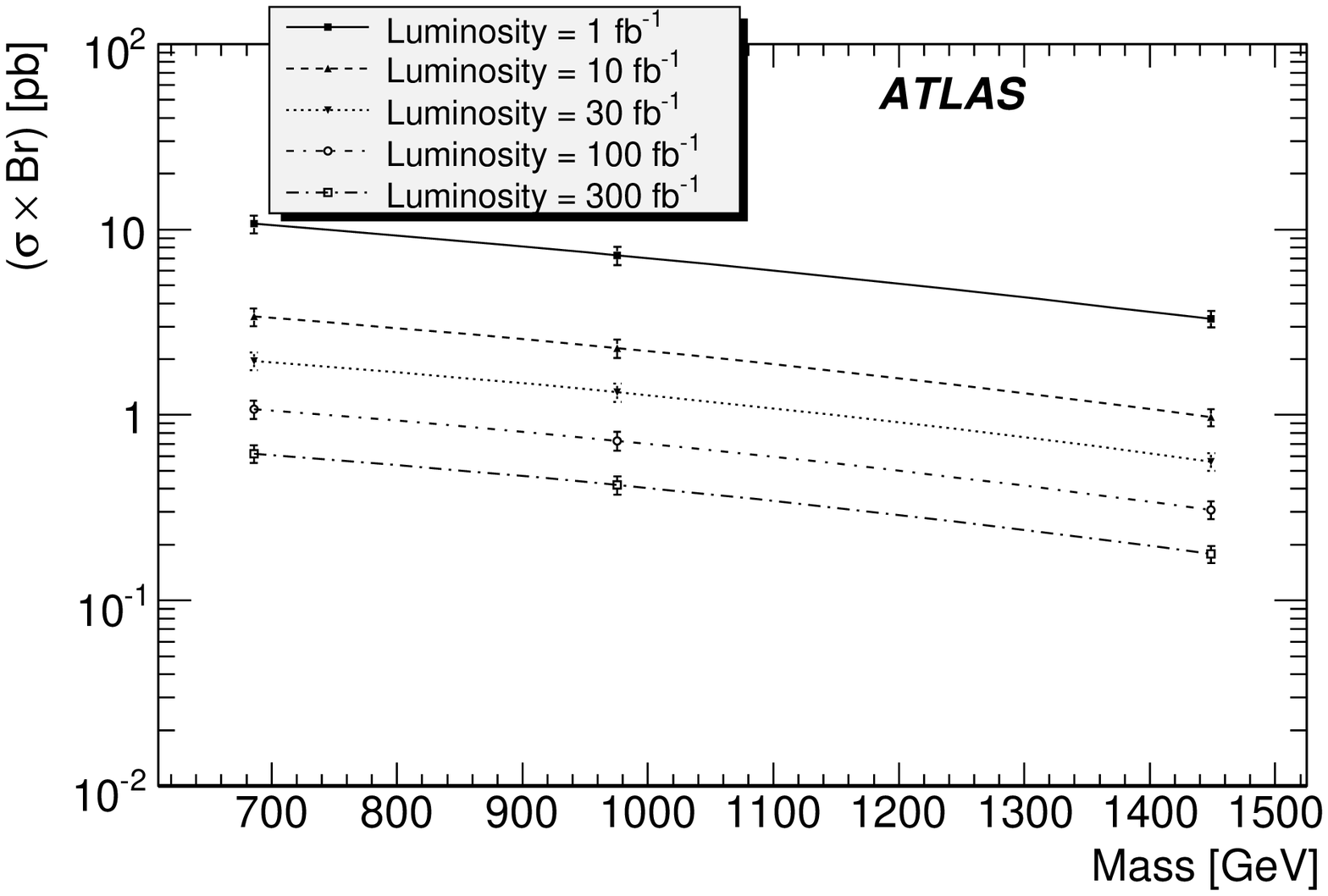}
  \caption{5$\sigma$ discovery potential of a generic narrow $t\bar{t}$ resonance as a function of the integrated luminosity.}
  \label{DiscovPot}
 \end{minipage}
\end{figure}

\subsection{ATLAS sensitivity to FCNC top quark decays}

We have studied the rare top quark decays via FCNC ($t\rightarrow qX$, $X=\gamma, Z, g$) using $t\bar{t}$ events in $1$~fb$^{-1}$ of simulated LHC data. One of the top quarks is assumed to decay through its dominant decay mode ($t\to bW$), while the other top quark decays via one of the FCNC modes ($t\to qZ$, $t\to q\gamma$, $t\to qg$). Due to the large QCD background, it is very difficult to search for FCNC signal using modes where $W$ or $Z$ decay hadronically. Due to this reason, only leptonic decays of both $W$ and $Z$ were taken into account. For signal events, we have used $t\bar{t} \to b\ell\nu qX$, where $X=\gamma,Z\to\ell\ell,g$ and $\ell=\mathrm{e},\mu$ and taken into account the expected Standard Model backgrounds.

 For $t\bar{t} \to bWq\gamma$, we require exactly one lepton with $p_{\mathrm{T}}>25$~GeV, at least two jets with $p_{\mathrm{T}}>20$~GeV, one $\gamma$ with $p_{\mathrm{T}}>25$~GeV and $\not\!p_{\mathrm{T}}>20$~GeV. For $t\bar{t}\to bWqg$, we require exactly one lepton with $p_{\mathrm{T}}>25$~GeV, exactly three jets with $p_{\mathrm{T}}>40, 20, 20$~GeV and $\not\!p_{\mathrm{T}}>20$~GeV. For $t\bar{t}\to bWqZ$, we require exactly three leptons with $p_{\mathrm{T}}>25, 15, 15$~GeV, at least two jets with $p_{\mathrm{T}}>30, 20$~GeV and $\not\!p_{\mathrm{T}}>20$~GeV. The neutrino four momentum was estimated using a kinematic fit[3]. The expected 95\%~ C.L. upper limits on the branching ratios for $t\to qZ, t\to q\gamma, t\to qg$ are 10$^{-3}$, 10$^{-3}$, 10$^{-2}$ respectively using $1$~fb$^{-1}$ simulated data. Figure 3 shows the expected 95\%~C.L. for the first $1$~fb$^{-1}$ in the absence of signal for the $t\to q\gamma$ and $t \to qZ$ channels.

\subsection{\textbf{$t\bar{t}$} resonances}

 The discovery potential for generic $t\bar{t}$ resonances with the ATLAS detector has been explored as a function of the resonance mass for the semileptonic $t\bar{t}$ channels[5]. $t\bar{t}$ resonances were produced with \textsc{Pythia} for $Z' \to t\bar{t}$ channel. The common selection criteria have been applied for event reconstruction. The main source of background for $t\bar{t}$ resonances is the Standard Model $t\bar{t}$ events (other backgrounds like $W$+jets are negligible). It is possible to discover a 700 GeV $Z'$ resonance produced with a $\sigma \times Br(Z' \to t\bar{t})$ of 11~pb with a 5$\sigma$ significance with 1~fb$^{-1}$ of data [Figure 4]. Using a model-independent approach, ATLAS can exclude Kaluza-Klein gluon resonances upto 1.5 TeV with only 1~fb$^{-1}$ data[3].

%\begin{thebibliography}{9}   % Use for  1-9  references
%\begin{thebibliography}{99} % Use for 10-99 references

%\bibitem{exampl-ref}

%\end{thebibliography}

\section{References}
%\begin{enumerate} 
 [1] F. Abe $et$ $al.$ (CDF Collaboration), Phys. Rev. Lett. \textbf{74}, 2626 (1995); S. Abachi $et$ $al.$ (D0 Collaboration, $ibid.$ \textbf{74}, 2632 (1995).
 
 [2] Lillie, B and Randall, L. and Wang, L.-T., hep-ph/0701166v1 (2007) 

 [3] ATLAS Collaboration, CERN-OPEN--2008-020, Geneva, 2008 to appear.

 [4] W. Bernreuther, Nucl. Phys. B \textbf{690} 81 (2004).

 [5] E. Cogneras and D. Pallin, ATL-PHYS-PUB-2006-033 (2006).

\section{Acknowledgements}
%\begin{acknowledgments}
%The authors wish to thank JACoW for their guidance in preparing this template.
The author would like to thank the organizers of ICHEP08 conference for creating fruitful collaborative environment. My sincere thanks to Antonio Onofre, Patrick Skubic and Martine Bosman for valuable suggestions.

%\end{acknowledgments}

\end{document}